\documentclass[prd,twocolumn,showpacs]{revtex4}
\usepackage{hyperref,amssymb,amsmath,mathrsfs,bm,graphicx}
\usepackage{enumitem}
\usepackage{color}
\begin{document}
\title{Interior solution for the Kerr metric} 
\author{J. L. Hernandez-Pastora}
\email{jlhp@usal.es}
\affiliation{Departamento de Matematica Aplicada and Instituto Universitario de
Fisica Fundamental y Matematicas, Universidad de Salamanca, Salamanca, Spain}
\author{L. Herrera}
\email{lherrera@usal.es}
\affiliation{Escuela de F\'\i sica, Facultad de Ciencias, Universidad Central de Venezuela, Caracas 1050, Venezuela and Instituto Universitario de F\'isica
Fundamental y Matem\'aticas, Universidad de Salamanca 37007, Salamanca, Spain}

\begin{abstract}
A, recently presented,  general procedure to find static and axially symmetric, interior solutions to the Einstein equations, is extended  to the stationary case, and applied  to find an interior solution for the Kerr metric.  The solution, which is  generated by an anisotropic fluid, verifies the energy conditions for a wide range of values of the parameters, and matches smoothly to  the Kerr solution, thereby representing a globally regular model describing a non spherical and rotating source of gravitational field. In the spherically symmetric limit, our model converges to the well known  incompressible perfect fluid solution.The key stone of our approach is based on an ansatz allowing to define the interior metric in terms of the exterior metric functions evaluated at the boundary source. The physical variables of the energy-momentum tensor are calculated explicitly, as well as the geometry of the source in terms of the relativistic multipole moments.
\end{abstract}
\date{\today}
\pacs{04.20.Cv, 04.20.Dw, 97.60.Lf, 04.80.Cc}
\maketitle

\section{Introduction}
Since the discovery of the Kerr  metric  \cite{Kerr} there have been many attempts to find a  physically meaningful matter distribution  that could serve as its source (see \cite{Her1, C, Wa, Her2, W, Kr, HM, HJ, H, DT, Pa, Via1, Via2, K, Az, K1} and references therein). However, no satisfactory solution has yet been found to this problem. 

It is the purpose of this work, to provide an interior solution to Einstein equations, satisfying all the usual physical conditions, and smoothly matched on the boundary surface of the fluid distribution, to the Kerr metric.
With this aim we shall  generalize a procedure, recently proposed to find sources of the Weyl metrics \cite{HHM}, to the stationary case.

As a particular example we shall find a source for the Kerr metric which consists in an anisotropic fluid,  satisfying the Darmois matching conditions on the boundary surface of the matter distribution, thereby excluding the presence of thin shells, and exhibiting a well behaviour of all physical variables, for a wide range of values of the parameters of the solution. These include values which are commonly assumed in realistic models of  rotating neutron stars and white dwarfs.

\section{The global model of a self-gravitating stationary source}

\subsection{The exterior metric}

The  line element for a vacuum stationary  and axially symmetric space--time, in Weyl canonical coordinates may be written as :
\begin{equation}
 ds^2_E=-e^{2\psi}(dt-w d\phi)^2+e^{-2\psi+2\Gamma}(d\rho^2+dz^2)+e^{-2\psi}\rho^2d\phi^2,
\label{1}
\end{equation}
where $\psi=\psi(\rho,z)$ ,  $\Gamma=\Gamma(\rho,z)$ and  $w=w(\rho,z)$ are functions of their arguments.

For vacuum space--times, Einstein's field equations imply for the
metric functions 
\begin{equation}f( f_{, \rho \rho}+\rho^{-1}
f_{, \rho}+f_{, zz})-f_{, \rho}^2-f_{, z}^2+\rho^{-2} f^4(w_{, \rho}^2+w_{, z}^2) = 0, \label{meq1}
\end{equation}
\begin{equation}f( w_{, \rho \rho}+\rho^{-1}
w_{, \rho}+w_{, zz})+2f_{, \rho}w_{, \rho}+2f_{, z}w_{, z} = 0, \label{meq2}
\end{equation}
with $f \equiv e^{2\psi}$ and
\begin{eqnarray}
\Gamma_{, \rho}&=&\frac 14 \rho f^{-2} (f_{, \rho}^2-f_{, z}^2)-\frac 14 \rho^{-1} f^{2} (w_{, \rho}^2-w_{, z}^2) \nonumber \\ 
\Gamma_{, z}&=& \frac 12 \rho f^{-2} f_{, \rho}f_{, z}-\frac 12 \rho^{-1} f^{2} w_{, \rho}w_{, z}. \label{meq3}
\end{eqnarray}

Notice  that  (\ref{meq1}),(\ref{meq2}) are  precisely
the integrability condition of (\ref{meq3}), that is: given any  $\psi$ and $w$ satisfying (\ref{meq1}),(\ref{meq2}),
a function $\Gamma$ satisfying (\ref{meq3}) always exists.

We can write the line element above, in  Erez-Rosen \cite{erroz}, or standard Schwarzschild--type coordinates $\{r,y\equiv \cos\theta\}$ or in  spheroidal prolate coordinates $\{x\equiv\frac{r-M}{M},y\}$ \cite{quev}:
\begin{equation}
\rho^2=r(r-2M)(1-y^2) \ , \quad z=(r-M)y,
\label{2}
\end{equation}
where $M$ is a constant which will be identified later.

In terms of the above coordinates the line element (\ref{1}) may be writen as:
\begin{widetext}
\begin{eqnarray}
 ds^2_E=-e^{2\psi(r,y)}(dt -w d\phi)^2+e^{-2\psi+2\left[\Gamma(r,y)-\Gamma^s
 \right]}dr^2+
 e^{-2\left[\psi-\psi^s\right]+
 2\left[\Gamma(r,y)-\Gamma^s\right]}r^2 d\theta^2
+ e^{-2\left[\psi-\psi^s\right]} r^2\sin^2\theta d\phi^2,
\label{exterior}
\end{eqnarray}
\end{widetext}

where  $\psi^s$ and $\Gamma^s$ are the metric functions corresponding to the Schwarzschild solution, namely,
\begin{equation}
\psi^s=\frac 12 \ln \left(\frac{r-2M}{r}\right) \, \quad \Gamma^s=-\frac 12 \ln \left[\frac{(r-M)^2-y^2M^2}{r(r-2M)}\right],
\label{schwfun}
\end{equation}
where the parameter $M$ is easily identified as the Schwarzschild mass.

\subsection{The interior metric}
We shall now assume for the interior axially symmetric line element:
\begin{eqnarray}
ds^2_I&=&-e^{2 \hat a} Z(r)^2 ( dt -\Omega d\phi)^2+\frac{e^{2\hat g-2\hat a}}{A(r)} dr^2+e^{2\hat g-2\hat a}r^2 d\theta^2\nonumber \\&+&e^{-2 \hat a}r^2 \sin^2\theta d\phi^2,
\label{interior}
\end{eqnarray}
with
\begin{equation}
\hat a \equiv a(r,\theta)-a^s(r) \ , \qquad  \hat g \equiv g(r,\theta)-g^s(r,\theta),
\label{funcint}
\end{equation}
 where $\Omega=\Omega(r,\theta)$, and $a^s(r)$ and  $g^s(r,\theta)$ are functions that, on the boundary surface, equal the metric functions corresponding to the Schwarzschild solution  (\ref{schwfun}), i.e.  $a^s(r_{\Sigma})=\psi^s_{\Sigma}$ and  $g^s(r_{\Sigma})=\Gamma^s_{\Sigma}$.  Also, $A(r)\equiv 1-pr^2$  and  $Z\equiv\displaystyle{ \frac 32 \sqrt{A(r_{\Sigma})}-\frac 12 \sqrt{ A(r)}}$, where $p$ is an arbitrary constant and the boundary surface of the source is defined by $r=r_{\Sigma}=const.$ 
 
The case $w=0$, $\hat g=\hat a=0$, corresponds to a spherically symmetric distribution, more specifically, to the well known incompressible (homogeneous energy density) perfect fluid sphere, and hence the matching of (\ref{interior}) with the Schwarzschild solution implies $p=\displaystyle{\frac{2M}{r_{\Sigma}^3}}$. The simple condition $w=0$ recovers, of course, the static case.

It should be noticed  that for  simplicity we consider here only matching surfaces of the form $r = r_\Sigma =
const$, of course more general surfaces with axial symmetry could however be considered as well.

\subsection{The matching conditions}

We shall now turn to the matching (Darmois) conditions \cite{26}. Thus the continuity of the first and the second fundamental form across the boundary surface implies the continuity of the metric functions and the continuity of the first derivatives  $\partial_r g_{tt}$, $\partial_r g_{\theta \theta}$, $\partial_r g_{\phi \phi}$, producing:
\begin{eqnarray}
&& a_{\Sigma}=\psi_{\Sigma} \ , \quad a^{\prime}_{\Sigma}=\psi^{\prime}_{\Sigma} \ , \quad g_{\Sigma}=\Gamma_{\Sigma} \ , \quad g^{\prime}_{\Sigma}=\Gamma^{\prime}_{\Sigma}, \nonumber\\ 
&&a^s_{\Sigma}=\psi^s_{\Sigma} \ , \quad (a^s)^{\prime}_{\Sigma}=(\psi^s)^{\prime}_{\Sigma}, \nonumber \\ && g^s_{\Sigma}=\Gamma^s_{\Sigma} \ , \quad (g^s)^{\prime}_{\Sigma}=(\Gamma^s)^{\prime}_{\Sigma},
 \nonumber\\
 && \Omega_{\Sigma}=w_{\Sigma} \ , \quad \Omega^{\prime}_{\Sigma}=w^{\prime}_{\Sigma},
\label{matchingcond}
\end{eqnarray}
where prime denotes partial derivative with respect to $r$  and subscript $\Sigma$ indicates that the quantity is evaluated on the boundary surface. It is important to keep in mind that we are using global coordinates $\{r,\theta\}$  on both sides of the boundary.

Thus, our line element (\ref{interior}) matches smoothly with any stationary exterior  (\ref{exterior}), provided conditions (\ref{matchingcond}) are satisfied.

In the particular case when we want to match our interior with the Schwarzschild exterior (the static limit), then $\psi=\psi^s$ and $\Gamma=\Gamma^s$, and the source is a perfect fluid  if  $\hat a=\hat g=\Omega=0$.

We shall now see how the field equations constrain further our possible interiors.

\subsection{The field equations and constraints}

Let us first  analyse the well known case when the interior is spherically symmetric, then   $\hat a=\hat g=0$, and the physical variables are obtained from the field equations for a perfect fluid, the result is well known and reads (in relativistic units)
\begin{eqnarray}
-T^0_0\equiv \mu&=&\frac{3p}{8 \pi},\nonumber\\
T^1_1=T^2_2=T^3_3\equiv P&=&\mu\left(\frac{\sqrt A-\sqrt{A_{\Sigma}}}{3 \sqrt{A_{\Sigma}}-\sqrt{A}}\right),
\label{eeesf}
\end{eqnarray}
with  $A=\displaystyle{1-\frac{2m(r)}{r}=1-p r^2=1-\frac{2M r^2}{r_{\Sigma}^3}}$,
where $\mu$ and $P$ denote the energy density and the isotropic pressure respectively, and  for the mass function $m(r)$ we have
\begin{equation}
m(r)=-4\pi\int^{r}_0r^2 T^0_0 dr,
\end{equation}
implying 
\begin{equation}
M\equiv m(r_{\Sigma})=-4\pi\int^{r_{\Sigma}}_0r^2 T^0_0 dr=\frac{p r_{\Sigma}^3}{2}.
\end{equation}

This model, which describes the well known incompressible perfect fluid sphere, is further restricted by the requirement that the pressure be regular and positive everywhere within the fluid distribution, which implies $\displaystyle{\tau\equiv{\frac{r_{\Sigma}}{M}}>\frac94}$. As it is evident from (\ref{eeesf}) the pressure vanishes at the boundary surface. 

Finally, if we impose the strong energy condition $P<\mu$, we should further restrict our model with the condition $\displaystyle{\tau>\frac83}$.

 We shall now proceed to consider the general, non--spherical case. Thus, for our line element  (\ref{interior}) we have the following non vanishing components of the energy--momentum tensor:
\begin{eqnarray}
-T^0_0&=&\kappa \left(8 \pi \mu+\hat p_{zz}-E+3 \delta  J_{+}+ \delta \Omega I\right),\nonumber\\
T^1_1&=& \kappa \left(8 \pi P-\hat p_{xx}-\delta J_{-}\right),\nonumber \\
T^2_2&=& \kappa \left(8 \pi P+\hat p_{xx}+\delta J_{-}\right),\nonumber \\
T^3_3&=&\kappa \left(8 \pi P-\hat p_{zz}+\delta \Omega I+\delta J_+\right),
\nonumber \\
T^3_0&=&-\kappa \delta I,
\label{eegeneral}
\end{eqnarray}
\begin{widetext}
\begin{equation}
T_1^2=g^{\theta\theta}T_{12}=-\frac{\kappa}{r^2}\left[2 {\hat a}_{,\theta}  \hat a^{\prime}-\hat g^{\prime}\frac{\cos\theta}{\sin\theta}-\frac{\hat g_{,\theta}}{r}+\frac{(1-A)}{r \sqrt A (3 \sqrt{A_{\Sigma}}- \sqrt{A})}(2 {\hat a}_{,\theta} -{\hat g}_{,\theta}  )-\delta \frac{\Omega_{,\theta}\Omega^{\prime}}{2 r^2\sin^2\theta}\right],
\label{pxy}
\end{equation}
\end{widetext}
with $\displaystyle{\kappa\equiv \frac{e^{2\hat a-2\hat g}}{8 \pi}}$, $\displaystyle{\delta\equiv e^{4\hat a }Z^2}$,  $\Omega_{,y}$ denotes derivative of  $\Omega$ with respect to $y\equiv \cos\theta$, and
\begin{eqnarray}
&&J_{\pm}=\left(\frac{-\Omega_{,y}}{2 r^2}\right)^2\pm A\left(\frac{\Omega^{\prime}}{2 r \sin\theta}\right)^2,\nonumber\\
&&I=\frac{\Omega^{\prime \prime}A}{2r^2 \sin^2\theta}  +\frac{2\Omega^{\prime}\hat a^{\prime}A}{r^2 \sin^2\theta}+\frac{\Omega_{,yy}}{2 r^4} +2 \frac{\Omega_{, y}\hat a_{,y}}{r^4}+\nonumber\\
&+&\frac{(1-A)\Omega^{\prime}}{2 r^3 \sin^2\theta}\left( \frac{4 \sqrt{A_{\Sigma}}-3\sqrt A}{3 \sqrt{A_{\Sigma}}-\sqrt A}\right),\nonumber\\
&&E=-2 \Delta \hat a+(1-A)\left[2 \frac{\hat a^{\prime}}{r}\frac{9 \sqrt{A_{\Sigma}}-4 \sqrt{A}}{3 \sqrt{A_{\Sigma}}- \sqrt{A}}+2 \hat a^{\prime \prime}-\hat g^{\prime \prime}\right],\nonumber\\
&&\Delta \hat a= \hat a^{\prime \prime}+2\frac{\hat a^{\prime}}{r}+\frac{{\hat a}_{,\theta \theta} }{r^2}+\frac{{\hat a}_{,\theta} }{r^2}\frac{\cos \theta}{\sin \theta},\nonumber \\
&&\hat p_{xx}=-\frac{{\hat a}_{,\theta} ^2}{r^2}-\frac{\hat g^{\prime}}{r}+\hat a^{\prime 2}+\frac{{\hat g}_{,\theta} }{r^2}\frac{\cos \theta}{\sin \theta}+\nonumber\\
&+&(1-A)\left[2 \frac{\hat a^{\prime}}{r}\frac{\sqrt{A}}{3 \sqrt{A_{\Sigma}}- \sqrt{A}}- \hat a^{\prime 2} +\frac{\hat g^{\prime}}{r}\frac{3 \sqrt{A_{\Sigma}}-2 \sqrt{A}}{3 \sqrt{A_{\Sigma}}-\sqrt A}\right], \nonumber \\
&&\hat p_{zz}=-\frac{{\hat a}^2_{,\theta} }{r^2}-\frac{\hat g^{\prime}}{r}-\hat a^{\prime 2}-\frac{{\hat g}_{,\theta \theta} }{r^2}-\hat g^{\prime \prime}+\nonumber\\
&+&(1-A)\left[-2 \frac{\hat a^{\prime}}{r}\frac{\sqrt{A}}{3 \sqrt{A_{\Sigma}}- \sqrt{A}}+ \hat a^{\prime 2} +2\frac{\hat g^{\prime}}{r}\right].
\label{eegeneraldet}
\end{eqnarray}

From the expressions above, using   (\ref{eegeneral})-(\ref{eegeneraldet}) and  introducing the dimensionless parameter $s\equiv r/r_{\Sigma}$, we can now obtain the explicit expressions for the physical variables, (with $A=1-(2s^2)/\tau$, $A_{\Sigma}=A(s=1)$):
\begin{widetext}
\begin{eqnarray}
-T^0_0&=&\frac{\kappa}{r_{\Sigma}^2}\left\lbrace\frac{6}{\tau}-\frac{{\hat a_{,\theta}}^2}{s^2}-A\left[ (\hat a_{,s})^2 +(\hat g_{,ss})-2(\hat a_{,ss})\right]
-\frac{{\hat g_{,\theta \theta}}}{s^2}+(1-2A)\frac{\hat g_{, s}}{s}
-  2(1-3A) \frac{\hat a_{, s}}{s}+\frac{2}{s^2}\left({\hat a_{,\theta \theta}}+{\hat a_{,\theta}}\frac{\cos\theta}{\sin\theta} \right)
\right\rbrace+\nonumber \\
&+&\frac{\kappa \delta}{r_{\Sigma}^4}\left\lbrace\frac 34 \left(\frac{(\Omega_{,y})^2}{s^4}+\frac{A(\Omega_{,s})^2}{s^2\sin^2\theta}\right)+\Omega\left[\frac{\Omega_{,ss}A}{2s^2 \sin^2\theta}  +\frac{2\Omega_{,s}\hat a_{,s} A}{s^2 \sin^2\theta}+\frac{\Omega_{,yy}}{2 s^4} +2\frac{\Omega_{, y}\hat a_{,y}}{s^4}+\frac{(1-A)\Omega_{,s}}{2 s^3 \sin^2\theta}\left( \frac{4 \sqrt{A_{\Sigma}}-3\sqrt A}{3 \sqrt{A_{\Sigma}}-\sqrt A}\right) \right] \right\rbrace,\nonumber\\
\label{tdens}
\end{eqnarray}
\end{widetext}

 \begin{widetext}
\begin{equation}
T^3_0=-\frac{\kappa \delta}{r_{\Sigma}^4}\left[\frac{\Omega_{,ss}A}{2s^2 \sin^2\theta}  +\frac{2\Omega_{,s}\hat a_{,s} A}{s^2 \sin^2\theta}+\frac{\Omega_{,yy}}{2 s^4} +2\frac{\Omega_{, y}\hat a_{,y}}{s^4}+\frac{(1-A)\Omega_{,s}}{2 s^3 \sin^2\theta}\left( \frac{4 \sqrt{A_{\Sigma}}-3\sqrt A}{3 \sqrt{A_{\Sigma}}-\sqrt A}\right) \right],
\label{t03}
\end{equation}
\end{widetext}

\begin{widetext}
\begin{eqnarray}
T_1^1&=&\frac{\kappa}{r_{\Sigma}^2}\left\lbrace\frac{6}{\tau}\frac{\sqrt A-\sqrt{A_{\Sigma}}}{3\sqrt{A_{\Sigma}}-\sqrt A}+\frac{{\hat a_{,\theta}}^2}{s^2}-(\hat a_{,s})^2 A-\frac{{\hat g_{,\theta}}}{s^2}\frac{\cos\theta}{\sin\theta}- 2 \frac{\hat a_{,s}}{s}\frac{\sqrt A (1-A)}{3\sqrt{A_{\Sigma}}-\sqrt A}-\frac{\hat g_{,s}}{s}\left[\frac{(1-A)(3\sqrt{A_{\Sigma}}-2\sqrt A)}{3\sqrt{A_{\Sigma}}-\sqrt A}-1 \right]\right\rbrace+\nonumber \\
&-&\frac{\kappa \delta}{r_{\Sigma}^4}\left\lbrace\frac 14 \left(\frac{(\Omega_{,y})^2}{s^4}-\frac{A(\Omega_{,s})^2}{s^2\sin^2\theta}\right)\right\rbrace,
\label{t11}
\end{eqnarray}
\end{widetext}

\begin{widetext}
\begin{eqnarray}
T_3^3&=&\frac{\kappa}{r_{\Sigma}^2}\left\lbrace\frac{6}{\tau}\frac{\sqrt A-\sqrt{A_{\Sigma}}}{3\sqrt{A_{\Sigma}}-\sqrt A}+\frac{{\hat a_{,\theta}}^2}{s^2}+(\hat a_{,s})^2 A+(\hat g_{,ss})+\frac{{\hat g_{,\theta \theta}}}{s^2}+ 2 \frac{\hat a_{,s}}{s}\frac{\sqrt A (1-A)}{3\sqrt{A_{\Sigma}}-\sqrt A}-\frac{\hat g_{,s}}{s} (1-2A) \right\rbrace+\nonumber \\
&+&\frac{\kappa \delta}{r_{\Sigma}^4}\left\lbrace\frac 14 \left(\frac{(\Omega_{,y})^2}{s^4}+\frac{A(\Omega_{,s})^2}{s^2\sin^2\theta}\right)+\Omega\left[\frac{\Omega_{,ss}A}{2s^2 \sin^2\theta}  +\frac{2\Omega_{,s}\hat a_{,s} A}{s^2 \sin^2\theta}+\frac{\Omega_{,yy}}{2 s^4} +2\frac{\Omega_{, y}\hat a_{,y}}{s^4}+\frac{(1-A)\Omega_{,s}}{2 s^3 \sin^2\theta}\left( \frac{4 \sqrt{A_{\Sigma}}-3\sqrt A}{3 \sqrt{A_{\Sigma}}-\sqrt A}\right) \right] \right\rbrace,\nonumber\\
\label{t33}
\end{eqnarray}
\end{widetext}

\begin{widetext}
\begin{equation}
T_1^2=-\frac{\kappa}{s^2r_{\Sigma}^3}\left[2 {\hat a}_{,\theta}  \hat a_{,s}-\hat g_{,s}\frac{\cos\theta}{\sin\theta}-\frac{\hat g_{,\theta}}{s}+\frac{2s (2 {\hat a}_{,\theta} -{\hat g}_{,\theta}  )}{ \sqrt{\tau-2s^2} (3 \sqrt{\tau-2}- \sqrt{\tau-2s^2})}\right]+\frac{\kappa \delta}{r_{\Sigma}^5}\left[\frac{\Omega_{,\theta}\Omega_{,s}}{2s^4 \sin^2\theta} \right].
\label{t12}
\end{equation}
\end{widetext}
 Obviously for any specific (non--spherical) model we need to provide explicit forms for  $\hat a$, $\hat g$ and $\Omega$, however even at this level of generality we can assure that the junction conditions  (\ref{matchingcond}) imply $(P_{rr}\equiv g_{rr}T^1_1)_{\Sigma}=0$. 

We shall first proceed to prove the above statement, and then we shall provide a general procedure to choose $\hat a$ and $\hat g$ producing physically meaningful  models.

It is always possible to choose the metric functions $\hat a$ and  $\hat g$ such that, once the junction conditions (\ref{matchingcond}) are satisfied, the angular derivatives of such functions are continuous, i.e.:  $({\hat a}_{,\theta})_{\Sigma}=({\psi}_{,\theta})_{\Sigma}$ and  $({\hat g}_{,\theta})_{\Sigma}=({\Gamma}_{,\theta})_{\Sigma}$, as well as $({\Omega}_{,\theta})_{\Sigma}=({w}_{,\theta})_{\Sigma}$ .

Then, using $A_{\Sigma}=\displaystyle{\frac{r_{\Sigma}-2M}{r_{\Sigma}}}$ in (\ref{eegeneral}), we obtain for  $\hat p_{xx}+\delta J_{-}$  on the boundary surface:
\begin{widetext}
\begin{eqnarray}
(\hat p_{xx}+\delta J_{-})_{\Sigma}&=&\frac{1}{r_{\Sigma}^2} \left[  -({\hat{\psi}}_{,\theta})_{\Sigma}^2 +({\hat{\Gamma}}_{,\theta})_{\Sigma} \frac{\cos\theta}{\sin\theta}+2M\hat{\psi}^{\prime}_{\Sigma}+
r_{\Sigma}(r_{\Sigma}-2M)\hat{\psi}^{\prime 2}_{\Sigma}\ -(r_{\Sigma}-M)\hat{\Gamma}^{\prime}_{\Sigma}+\right.\nonumber\\
&+&\left.\frac{r_{\Sigma} e^{4\psi}}{4(r_{\Sigma}-2M)\sin^2\theta}\left(\frac{\dot{w_{\Sigma}}^2}{r_{\Sigma}^2}-(w_{\Sigma}^{\prime})^2\frac{r_{\Sigma}-2M}{r_{\Sigma}}\right)\right],
\label{pxxe}
\end{eqnarray}
\end{widetext}
where $\hat{\psi}_{\Sigma}\equiv \psi_{\Sigma}-\psi^s_{\Sigma}$,  $\hat{\Gamma}_{\Sigma}\equiv \Gamma_{\Sigma}-\Gamma^s_{\Sigma}$, and  dot  also denotes partial derivative with respect to $\theta$.

Taking into account   the Einstein's vacuum equations $(G_{\alpha \beta}=0)$, we find the following relation for the derivatives of the metric function $w$, from the Einstein  tensor  component $G_{11}=0$:
\begin{widetext}
\begin{eqnarray}
\frac{-r e^{4\psi}}{4(r-2M)\sin^2\theta}\left[\frac{\dot{w}^2}{r^2}-(w^{\prime})^2\frac{r-2M}{r}\right]&=&
-({\psi}_{,\theta})^2 +({\hat{\Gamma}}_{,\theta}) \frac{\cos\theta}{\sin\theta}+
r(r-2M)\psi^{\prime 2}-\frac{M^2}{r(r-2M)}\ -(r-M)\hat{\Gamma}^{\prime},
\end{eqnarray}
\end{widetext}
implying the vanishing of $(T_1^1)_{\Sigma}$.

In a similar way it can be shown that  $T_1^2$ vanishes on the boundary surface, if we take into account that the Einstein vacuum equation  $G_{12}=0$ produces
\begin{widetext}
\begin{equation}
\frac{w^{\prime}\dot{w}}{\sin^2\theta}
e^{4\psi}=4 r(r-2M)\dot{\psi}\psi^{\prime}-2(r-M)\dot{\hat{\Gamma}}-r(r-2M)2\hat{\Gamma}^{\prime}\frac{\cos\theta}{\sin\theta},
\end{equation}
\end{widetext}
and from (\ref{pxy}), we have the following expression for $T_1^2$ on the boundary
\begin{eqnarray}
(T_1^2)_{\Sigma}&=&2(\hat{\psi}_{, \theta})_{\Sigma}\hat{ \psi}^{\prime}_{\Sigma}-\hat{\Gamma}^{\prime}_{\Sigma}\frac{\cos\theta}{\sin\theta}+\nonumber \\
&-&\frac{r_{\Sigma}-M}{r_{\Sigma}(r_{\Sigma}-2M)}(\hat{\Gamma}_{, \theta})_{\Sigma}+\frac{2M}{r_{\Sigma}(r_{\Sigma}-2M)}(\psi_{, \theta})_{\Sigma}+\nonumber\\
&-&\frac{w^{\prime}_{\Sigma}(w_{, \theta})_{\Sigma}e^{4\psi_{\Sigma}}}{2 r_{\Sigma}(r_{\Sigma}-2M) \sin^2\theta}.
\end{eqnarray}

This last condition $(T_1^2)_{\Sigma}=0$, which follows from the Darmois conditions,  and therefore  is necessary, in order to avoid the presence of shells on the boundary surface, can be obtained  at once from a simple inspection of the equation (22) in \cite{H1}.

\subsection{The ansatz for the metric functions}

We shall provide a general procedure to choose $\hat a$, $\hat g$ and $\Omega$ producing physically meaningful  models. With this aim,  besides the fulfillment of the junction conditions (\ref{matchingcond}), we shall require that all physical variables be regular within the fluid distribution and the energy density to be positive.

To ensure the fulfillment of the junction conditions  (\ref{matchingcond}), we may write without loss of generality, 
\begin{eqnarray}
\hat a&=&\hat \psi_{\Sigma}(\theta)+\hat \psi^{\prime}_{\Sigma}(\theta) (r-r_{\Sigma})+\Lambda(r,\theta)(r-r_{\Sigma})^2\nonumber\\
\hat g&=&\hat \Gamma_{\Sigma}(\theta)+\hat \Gamma^{\prime}_{\Sigma}(\theta) (r-r_{\Sigma})+\Xi(r,\theta)(r-r_{\Sigma})^2, \nonumber \\
\Omega&=&w_{\Sigma}(\theta)+w^{\prime}_{\Sigma}(\theta) (r-r_{\Sigma})+\Pi(r,\theta)(r-r_{\Sigma})^2,
\label{ayg}
\end{eqnarray}
where $ \Lambda(r,\theta)$, $\Xi(r,\theta)$ and $\Pi(r,\theta)$ are so far two arbitrary functions of their arguments.

On the other hand, to guarantee a good behaviour of the physical variables at the center of the distribution we shall demand: 
\begin{eqnarray}
&\hat a^{\prime}_0={\hat a}_{,\theta 0}={\hat a }_{,\theta  \theta 0}={\hat a}^{\prime}_{,\theta 0}={\hat a }^{\prime}_{,\theta \theta 0}=0 \nonumber \\
&\hat g^{\prime}_0={\hat g}_{,\theta 0}={\hat g }_{,\theta \theta 0}={\hat g}^{\prime}_{,\theta 0}={\hat g }^{\prime}_{,\theta \theta 0}=0 \nonumber\\
&\hat g^{\prime \prime}_0={\hat g}^{\prime \prime}_{,\theta 0}=0,
\label{condcero}
\end{eqnarray}
\begin{eqnarray}
&\Omega_0=\Omega^{\prime}_0=\Omega^{\prime \prime}_0=\Omega^{\prime \prime \prime}_0=0 \nonumber \\
&\dot{\Omega}_0=\dot{\Omega}^{\prime}_0=0 \nonumber\\
&\ddot{\Omega}_0=\ddot{\Omega}^{\prime}_0=
\ddot{\Omega}^{\prime \prime}_0=\ddot{\Omega}^{\prime \prime \prime}_0=0,
\label{condceroomega}
\end{eqnarray}
where (\ref{pxy}, \ref{eegeneraldet}) have been used, and the subscript $0$ indicates that the quantity is evaluated at the center of the distribution. 

Using the conditions above in  (\ref{ayg}) we may write for  $\Lambda$, $\Xi$ and $\Pi$
\begin{eqnarray}
\Lambda(r,\theta)&=&\Lambda_0(\theta)+\Lambda^{\prime}_0(\theta) r+F(r,\theta)\nonumber \\
\Xi(r,\theta)&=&\Xi_0(\theta)+\Xi^{\prime}_0(\theta) r+\Xi^{\prime \prime}_0(\theta) r^2 +G(r,\theta)\nonumber\\
\Pi(r,\theta)&=&\Pi_0(\theta)+\Pi^{\prime}_0(\theta) r+\Pi^{\prime \prime}_0(\theta) r^2 +\nonumber\\
&+&\Pi^{\prime \prime \prime}_0(\theta) r^3 +H(r,\theta)
\end{eqnarray}
with  $F(0,\theta)=F^{\prime}(0,\theta)=0$ and  $G(0,\theta)=G^{\prime}(0,\theta)=G^{\prime \prime}(0,\theta)=0$, as well as $H(0,\theta)=H^{\prime}(0,\theta)=H^{\prime \prime}(0,\theta)=H^{\prime \prime \prime}(0,\theta)=0$.
Then we can finally write for $\hat a$ and $\hat g$ (please notice a that there is a misprint in Eq.(43) in \cite{HHM}, there it should read $\Gamma$ instead of $\psi$ in the two terms within the round brackets for the metric function $\tilde g(r,\theta$))
\begin{eqnarray}
\hat a(r,\theta)&=&r^2\left(-\frac{\hat \psi^{\prime}_{\Sigma}}{r_{\Sigma}}+3\frac{\hat \psi_{\Sigma}}{r_{\Sigma}^2}\right)+r^3 \left(-\frac{\hat \psi^{\prime}_{\Sigma}}{r_{\Sigma}^2}-2\frac{\hat \psi_{\Sigma}}{r_{\Sigma}^3}\right)\nonumber \\&+&(r-r_{\Sigma})^2F(r,\theta)
\label{aygdefa}
\end{eqnarray}

\begin{eqnarray}
\hat g(r,\theta)&=&r^4\left(\frac{\hat \Gamma^{\prime}_{\Sigma}}{r_{\Sigma}^2}-3\frac{\hat \psi_{\Sigma}}{r_{\Sigma}^3}\right)+r^3 \left(-\frac{\hat \Gamma^{\prime}_{\Sigma}}{r_{\Sigma}^2}+4\frac{\hat \psi_{\Sigma}}{r_{\Sigma}^3}\right)\nonumber \\&+&(r-r_{\Sigma})^2G(r,\theta).
\label{aygdef}
\end{eqnarray}

\begin{eqnarray}
\Omega(r,\theta)&=&r^4\left(\frac{-w^{\prime}_{\Sigma}}{r_{\Sigma}^3}+5\frac{w_{\Sigma}}{r_{\Sigma}^4}\right)+r^5 \left(\frac{w^{\prime}_{\Sigma}}{r_{\Sigma}^4}-4\frac{w_{\Sigma}}{r_{\Sigma}^5}\right)\nonumber \\&+&(r-r_{\Sigma})^2H(r,\theta).
\label{omdef}
\end{eqnarray}

The metric functions obtained so far, satisfy the junction conditions  (\ref{matchingcond}) and produce physical variables which are regular within the fluid distribution.  Furthermore  the vanishing of  $\hat g$ on the axis of symmetry, as required by the regularity conditions,  necessary to ensure elementary flatness in the vicinity of  the axis of symmetry, and in particular at the center (see \cite{1n}, \cite{2n}, \cite{3n}),
 is assured by the fact that $\hat \Gamma_{\Sigma}$ and $\hat \Gamma^{\prime}_{\Sigma}$ vanish on the axis of symmetry. 
Furthermore, the good behaviour of the function $\Omega$ on the symmetry  axis is fulfilled since $w_{\Sigma}$ and $w_{\Sigma}^{\prime}$ vanish when $y=\pm 1$.
 Finally, let us note that the energy-momentum tensor components (\ref{eegeneral}), (\ref{pxy}) do not diverge on the symmetry axis because the first  derivative with respect to the angular variable $\theta$ of  both  $\Omega^{\prime}$ and $\Omega^{\prime \prime}$ vanishes on the symmetry axis since not only $w_{\Sigma}$ and $w_{\Sigma}^{\prime}$ vanish there, but their first derivatives with respect to $\theta$ vanish as well.

So far we have presented the general procedure to build up sources for any stationary metric, in what follows, we shall illustrate the method with the example of Kerr metric.

\section{Particular solution}

\subsection{A source for the exterior Kerr's solution}

The Kerr metric in Weyl  coordinates is given by the following metric functions
\begin{equation}
f=\frac{(r_1+r_2)^2(1-j^2)-4 M^2(1-j^2)+j^2(r_1-r_2)^2}{(r_1+r_2+2M)^2(1-j^2)+j^2(r_1-r_2)^2},
\label{fkerr}
\end{equation}
\begin{equation}
e^{2\Gamma}=\frac{(r_1+r_2)^2(1-j^2)-4 M^2(1-j^2)+j^2(r_1-r_2)^2}{4r_1r_2(1-j^2)},
\label{gakerr}
\end{equation}
\begin{equation}
w=\frac{j(2M+r_1+r_2)(4M^2(1-j^2)-(r_1-r_2)^2)}{(r_1+r_2)^2(1-j^2)-4 M^2(1-j^2)+j^2(r_1-r_2)^2},
\label{omkerr}
\end{equation}
where $j\equiv \frac{J}{M^2}=a/M$ denotes the dimensionless parameter representing the angular momentum of the source, and  is related  to the rotation parameter $a$ of Kerr in its well-known Boyer-Lindquist representation. Also, $r_{1,2}\equiv \displaystyle{\sqrt{\rho^2+
(z\pm M\sqrt{1-j^2})^2}}$, which in the Erez-Rosen coordinates becomes
\begin{equation}
r_{1,2}^2=\left[(r-M)\pm My(r-M)\sqrt{1-j^2}\right]^2-M^2 j^2(1-y^2).
\end{equation}

  As is well known, the relativistic multipole moments (RMM) \cite{geroch, ger, ger3, han, th},  of the Kerr solution, are easily given in terms of the rotation parameter $j$ (as well  as $a$). In fact, using  the FHP method \cite{fhp}  in order to calculate the RMM, these are expressed  in terms of the expansion coefficients ($m_k$)  of the Ernst potential on the axis of symmetry, but the Kerr solution is the only one verifying that each  RMM ($M_k$) at any order $k$ is just equal to the corresponding coefficient $m_k$ for such order, which implies,
\begin{equation}
m_k=M_k=M (i a)^k
\end{equation}
Therefore, the massive RMM (even orders)  and the rotational RMM (odd orders) can be expressed as follows
\begin{equation}
M_{2l}=(-1)^l M^{2l+1} j^{2l} \quad , \quad M_{2l+1}=i (-1)^l M^{2l+2} j^{2l+1}.
\label{RMMkerr}
\end{equation}

In particular let us remind that a first conclusion derived from these RMM  (\ref{RMMkerr}), is that the rotation of the object leads to a negative quadrupole massive moment, $ {\displaystyle q\equiv \frac{M_2}{M^3}=- j^2}$, i.e. all the possible sources of Kerr solution are oblate.

Let us next, consider the line element  (\ref{interior}) with   metric functions given by (\ref{aygdefa}) and (\ref{aygdef})
\begin{eqnarray}
\hat a(r,\theta)&=&\hat \psi_{\Sigma} s^2(3-2s)   +r_{\Sigma}\hat \psi^{\prime}_{\Sigma}s^2(s-1),\nonumber \\
\hat g(r,\theta)&=&\hat \Gamma_{\Sigma} s^3(4-3s)   +r_{\Sigma}\hat \Gamma^{\prime}_{\Sigma}s^3(s-1),
\nonumber \\
\Omega(r,\theta)&=&w_{\Sigma} s^4(5-4s)+r_{\Sigma}w_{\Sigma}^{\prime} s^4(s-1).
\label{aygyomsimple}
\end{eqnarray}
with $s\equiv r/r_{\Sigma} \in \left[0,1\right]$.

For the Kerr  solution we may write:
\begin{widetext}
\begin{eqnarray}
&&\hat \psi_{\Sigma}\equiv \psi_{\Sigma}-\psi^s_{\Sigma}=\frac 12\ln\left\lbrace\frac{\tau}{\tau-2}\frac{N+ r_1^{\Sigma}r_2^{\Sigma} (2j^2-1)}{N+ r_1^{\Sigma}r_2^{\Sigma} (2j^2-1)-2(1-j^2)(r_1^{\Sigma}+r_2^{\Sigma}+2)} \right\rbrace,\nonumber\\
&&\hat \Gamma_{\Sigma}\equiv \Gamma_{\Sigma}-\Gamma^s_{\Sigma}=\frac 12\ln\left\lbrace\frac{(\tau-1)^2-y^2}{\tau(\tau-2)}\frac{N+ r_1^{\Sigma}r_2^{\Sigma} (2j^2-1)}{2 r_1^{\Sigma}r_2^{\Sigma}(j^2-1)} \right\rbrace, \nonumber\\
&&w_{\Sigma}=M j \frac{(N+ r_1^{\Sigma}r_2^{\Sigma}) (2+r_1^{\Sigma}+r_2^{\Sigma})}{(j^2-1)\left[(j^2-1)(N+ r_1^{\Sigma}r_2^{\Sigma})+2(N+j^4)\right]},
\label{prepsimq1}
\end{eqnarray}
\end{widetext}
with 
\begin{equation}
r_{1,2}^{\Sigma}=\sqrt{\left(\tau -1\pm y\sqrt{1-j^2}\right)^2-j^2 (1-y^2)},
\end{equation}
\begin{equation}
N\equiv -\tau (\tau-2)+(1-y^2)-j^2.
\end{equation}

A straightforward calculation, using (\ref{tdens})--(\ref{t12}) allows us to find the explicit expressions for the physical variables, these are displayed in figures (3)--(7). However, before entering into a detailed discussion of these figures, 
we shall  carry out some calculations with the purpose of providing some information about the ``shape'' of the source. In particular we shall see how it is related with the rotation parameter $j$ of the source. 

For doing so, using our interior metric, we shall calculate the proper length  $l_z$ of the object  along the axis  $z$, and the proper equatorial radius $l_{\rho}$:
\begin{equation}
l_z\equiv\int_0^{r_{\Sigma}}\frac{e^{\hat g(y=1)-\hat a(y=1)}}{\sqrt A} dz \ , \  l_{\rho}\equiv\int_0^{r_{\Sigma}}\frac{e^{\hat g(y=0)-\hat a(y=0)}}{\sqrt A} d\rho,
\end{equation}
where ${\rho,z}$ are the cylindrical coordinates associated to the  Erez-Rosen coordinates.

Obviously in the spherical case ($\hat a=\hat g=\Omega=0$), both lengths are identical:
\begin{equation}
l_z^s=l_{\rho}^s=\int_0^{r_{\Sigma}}\frac{d\xi}{\sqrt{1-p\xi^2}} =r_{\Sigma}\sqrt{\frac{\tau}{2}} \arcsin\sqrt{\frac{2}{\tau}},
\label{lzrho}
\end{equation}
where the fact that  $p=\displaystyle{\frac{2}{\tau r_{\Sigma}^2}}$ has been taken into account and where  $l_z^s$, $l_{\rho}^s$ denote the lengths corresponding to the spherical case. 

 It would be convenient to introduce here the concept of ellipticity ($e$), which in terms of  $l_z$ and  $l_{\rho}$, is defined as $e\equiv 1-\frac{l_{\rho}}{l_z}$. The two extreme values of this parameter are $e=0$, which corresponds to a spherical object, and   $e=1$ for the limiting  case when  the source is represented by a disk. In between of these two extremes we have  $e>0$ for a  prolate source and  $e<0$ for an oblate one.

In the general (non--spherical case) we must compare function $\displaystyle{e^{-\hat a(y=1)}}$ with  $\displaystyle{e^{\hat g(y=0)-\hat a(y=0)}}$, since $\hat g(y=\pm 1)$ vanishes along the axis.
  It can be seen that the sign of both  functions is positive, and their relative magnitudes verify the inequality $\displaystyle{e^{-\hat a(y=1)}} > e^{\hat g(y=0)-\hat a(y=0)}$ for all values of $s$ in the range $s\in [0,1]$, no matter the sign of the  parameter   $j$,  leading to  the well  known result that the rotation of the object always generates an oblate source ($q<0$) since $l_z<l_{\rho}$. In the figure 1, one example is shown.

 \begin{figure}[h]
\includegraphics[scale=0.21]{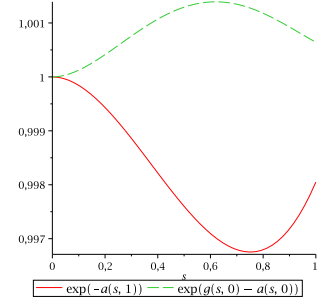}
\caption{\label{longitudes} {\it  Functions  $exp(\hat g(s,y=0)-\hat a(s,y=0))$ and  $exp(-\hat a(s,y=1))$  for  values of the rotation parameter $j=\pm 0.1$ and   $\tau=2.7$.}}
\end{figure}

Figure 2 shows the ellipticity $e$  of the source as a function  of the rotation parameter $j$, for different values of the  parameter $\tau$. As can be seen, the relation between $e$ and $j$ for any value of $\tau$ shows that  the  greater  is $j$,   the greater  is $e$, and therefore the shape of the source is more oblate. Of course, for $j=0$ (static case) we recover the sphericity ($e=0$). It is also observed from the figure 2 that the deformation of the source with respect to the spherical case, for any fixed value of $j$, is smaller for larger values of  $\tau$ (less compact object).

\begin{figure}[h]
 \includegraphics[scale=0.31]{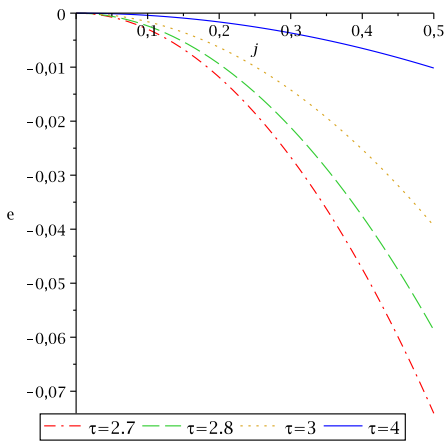}  
\caption{\label{ellip} {\it  Relation between the ellipticity of the source $e$ and its  rotation parameter $j$ for different values of $\tau$.}}
\end{figure}

Let us now turn back to the physical variables of our model. Figure  3 exhibits the behaviour of the radial pressure $P_{rr}\equiv g_{rr} T_1^1$  for different values of  $j$.

In it, we observe the variation of the radial pressure with respect to the spherically symmetric case  ($j=0$). This variation is smaller for angle values close to the equator, as it is apparent for $y=0.3$. Notice that the radial pressure is positive, with negative pressure gradient, and vanishes on the boundary surface.

\begin{figure}[h]
$$
\begin{array}{cc}
 \includegraphics[scale=0.21]{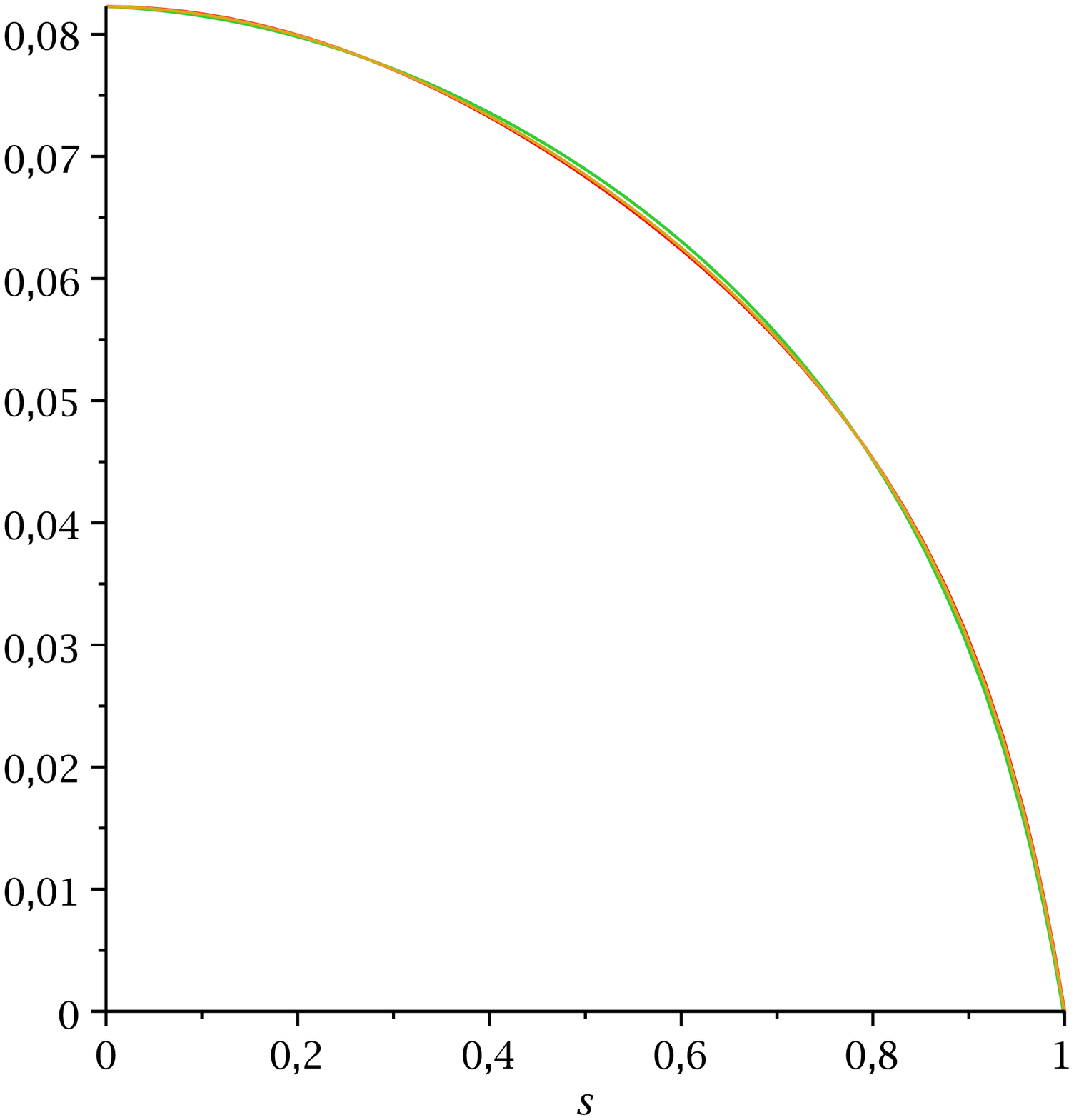}& \includegraphics[scale=0.21]{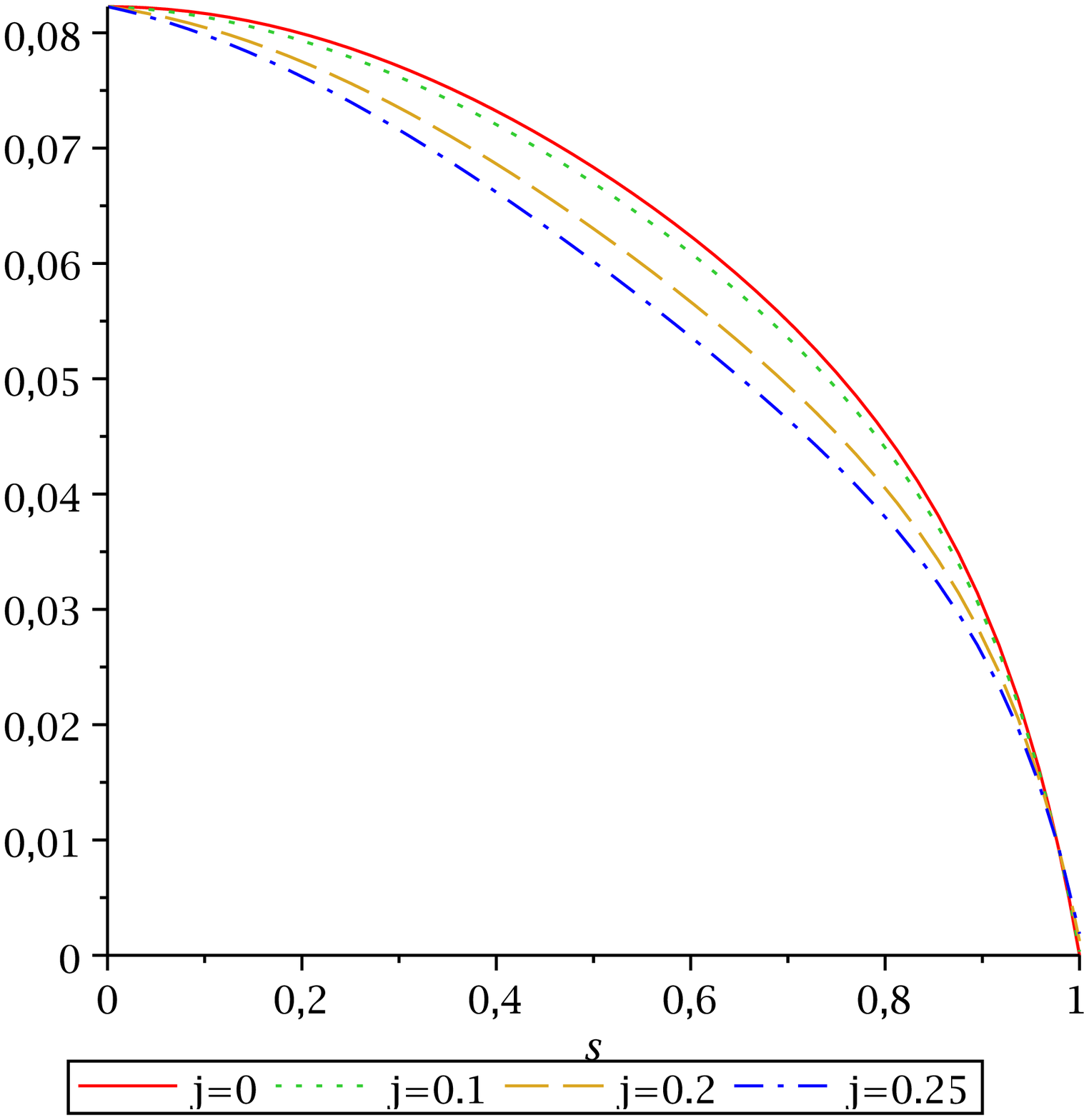}  \nonumber\\
(a) & (b) \nonumber
\end{array}
$$
\caption{\label{prr} {\it  Four profiles of  $r_{\Sigma}^2P_{rr}\equiv r_{\Sigma}^2g_{rr} T_1^1$, as function of $s$,   for  $y=0.3$ (graphic a), and  $y=1$ (graphic b) with  $\tau=2.7$, and   different values of $j$.}}
\end{figure}

Figures  4 and  5  depict the behaviour  of different energy momentum components, for a specific choice of the parameters $q$ and $\tau$. 
\begin{figure}[h]
$$
\begin{array}{cc}
 \includegraphics[scale=0.21]{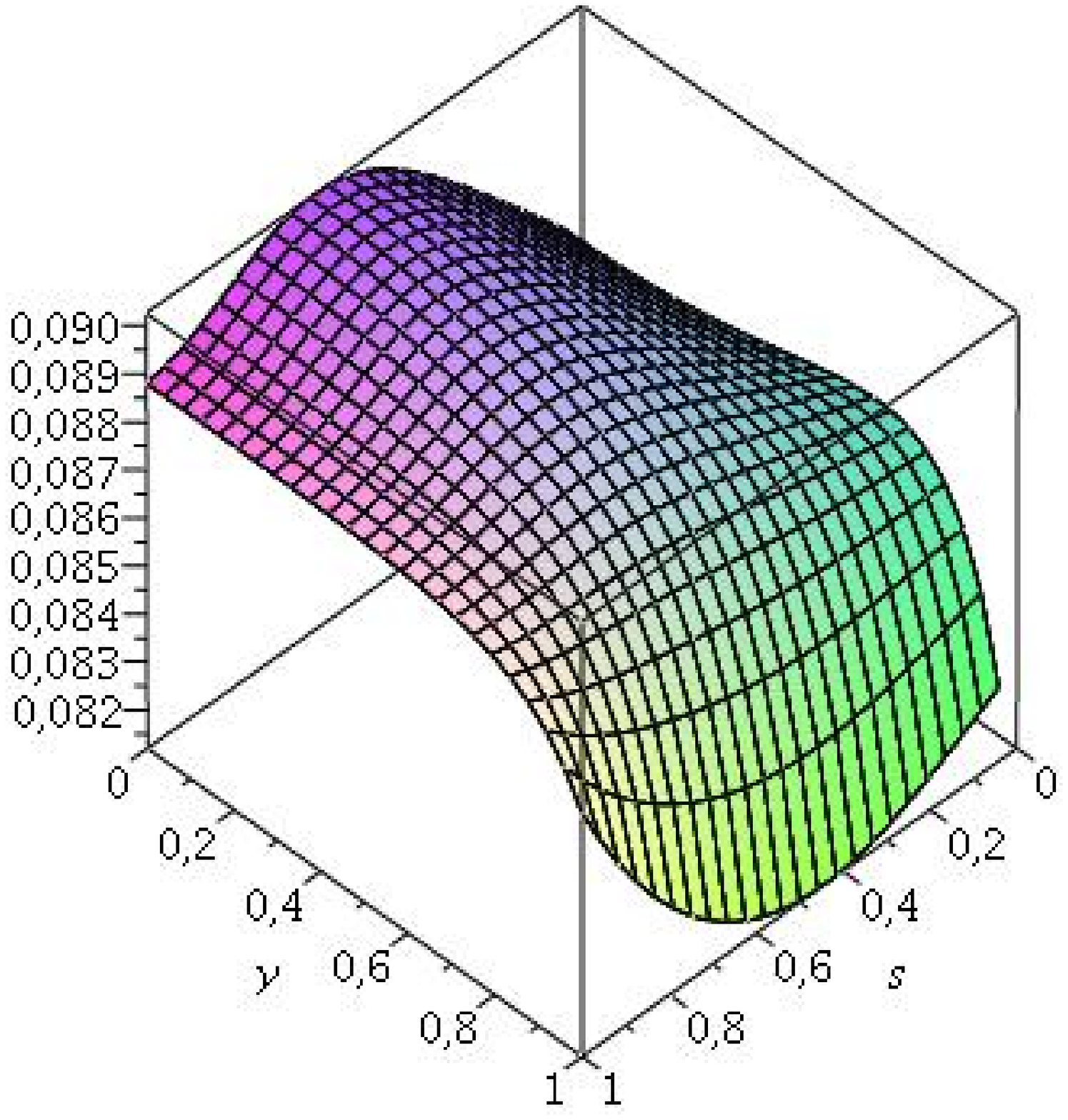}& \includegraphics[scale=0.21]{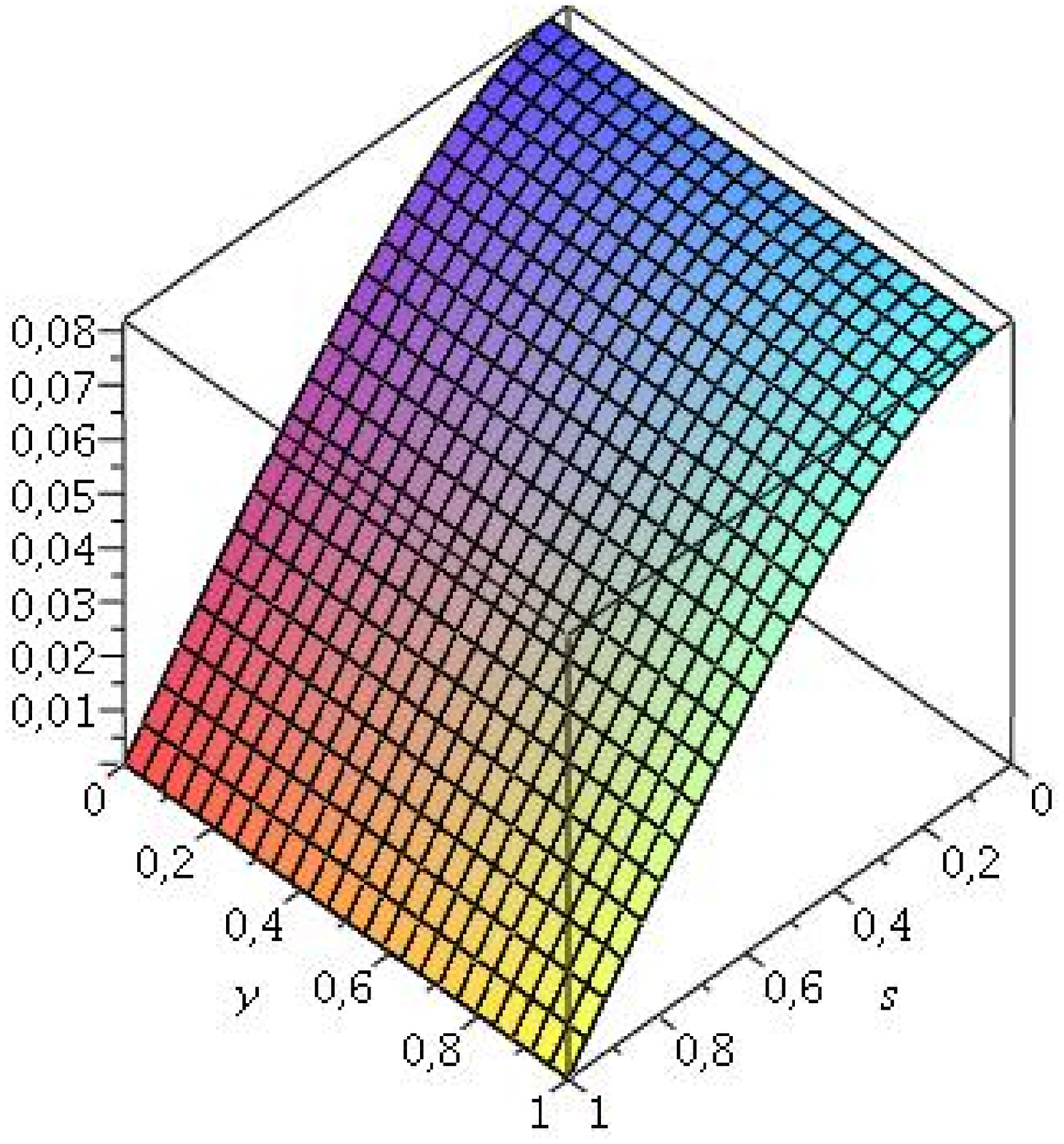}  \nonumber\\
(a) & (b) \nonumber
\end{array}
$$
\caption{\label{Ts} {\it   $-r_{\Sigma}^2T_0^0$ (graphic a), and $r_{\Sigma}^2T_1^1$  (graphic b), as functions  of  $y=\cos\theta$  and  $s$, with  $j=0.1$ and  $\tau=2.7$.}}
\end{figure}

\begin{figure}[h]
$$
\begin{array}{cc}
 \includegraphics[scale=0.21]{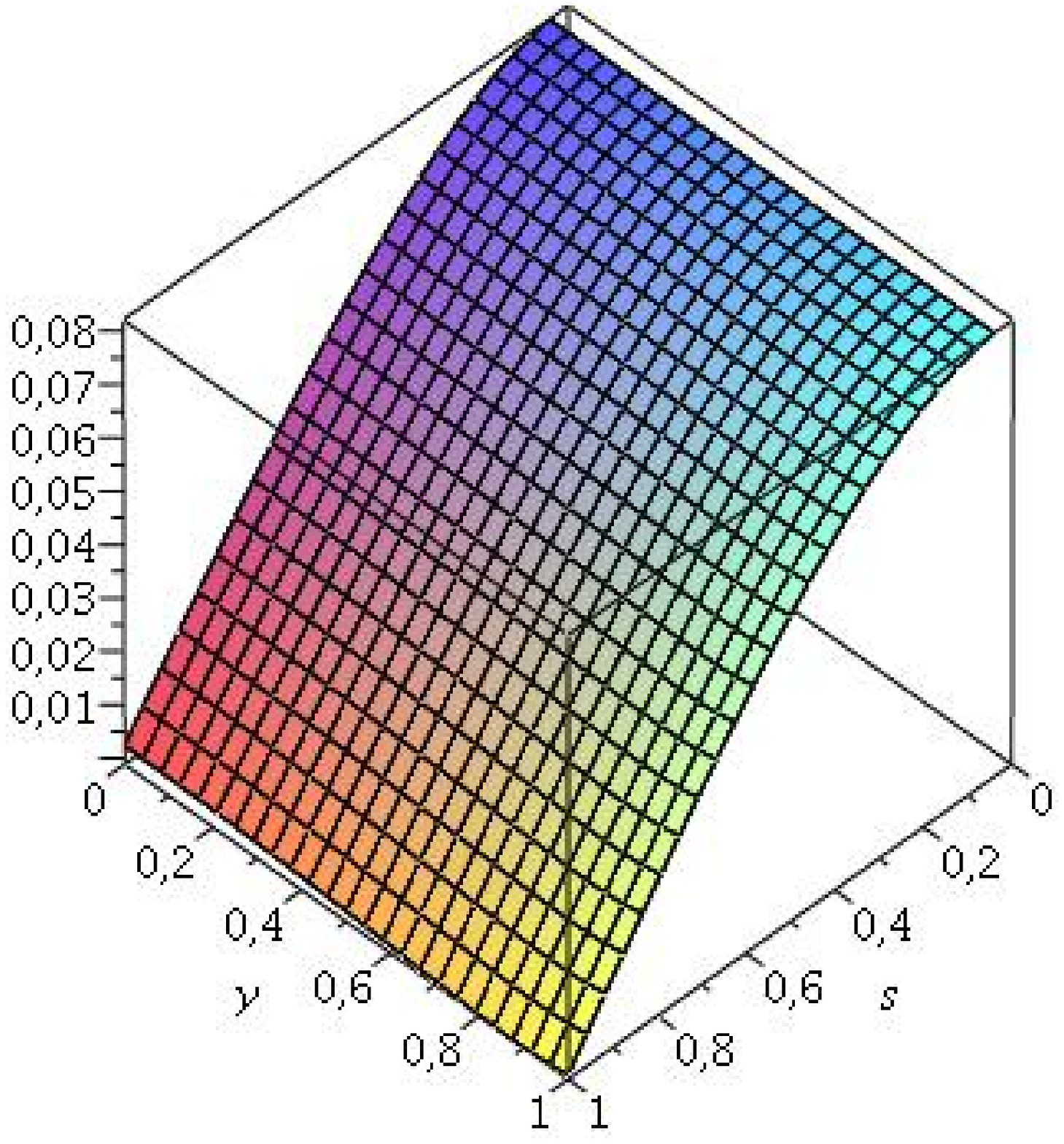}& \includegraphics[scale=0.21]{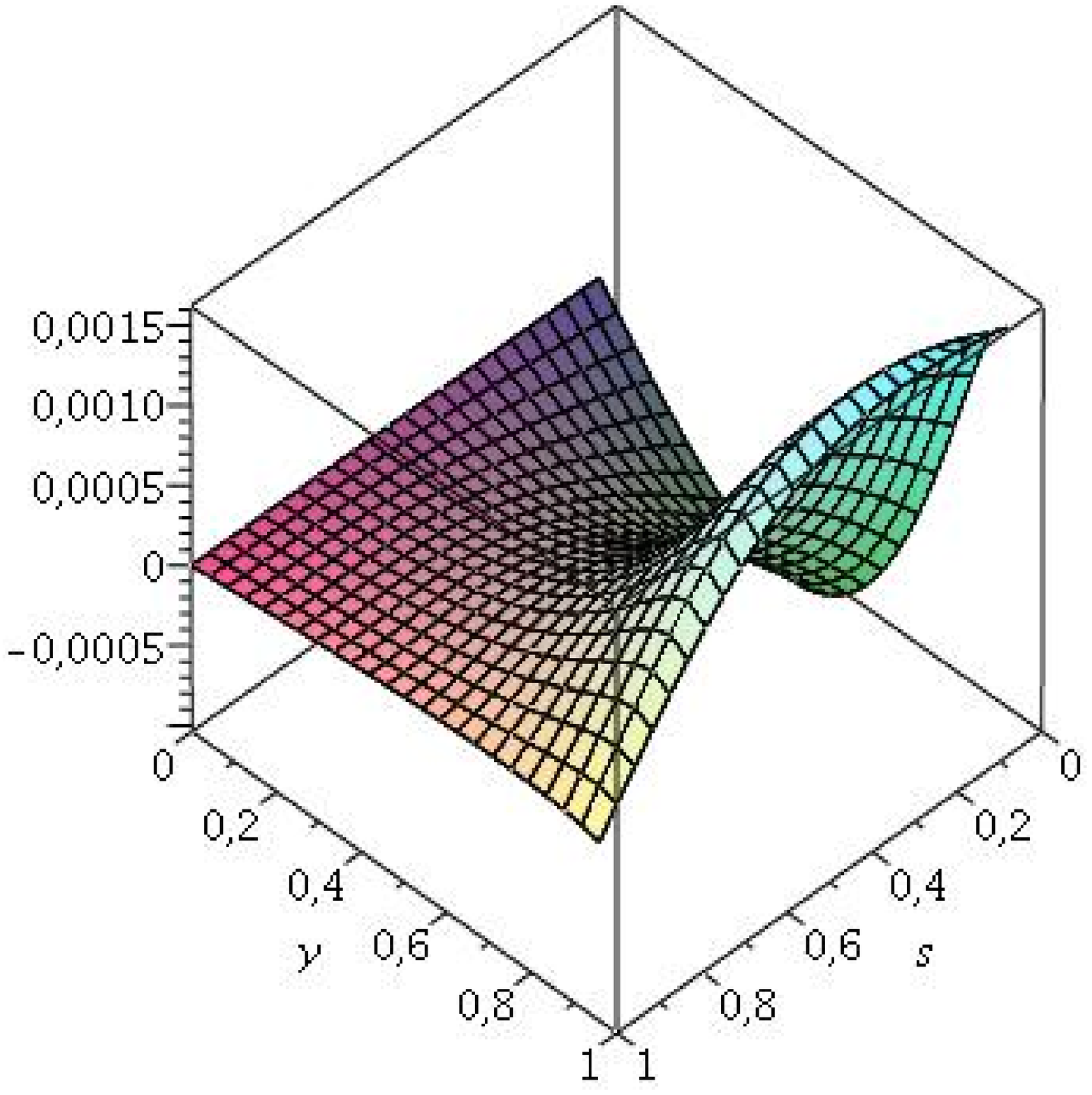}  \nonumber\\
(a) & (b) \nonumber
\end{array}
$$
\caption{\label{Ts12} {\it   (a)  $r_{\Sigma}^2T_3^3$,and (b) $r_{\Sigma}^3T_1^2$ as functions of $y$ and $s$.}}
\end{figure}

Figure 6, shows the verification of the strong energy condition $(- T_0^0)-T_1^1>0$ (also  for a specific choice of the parameters $j$ and $\tau$). 
\begin{figure}[h]
 \includegraphics[scale=0.31]{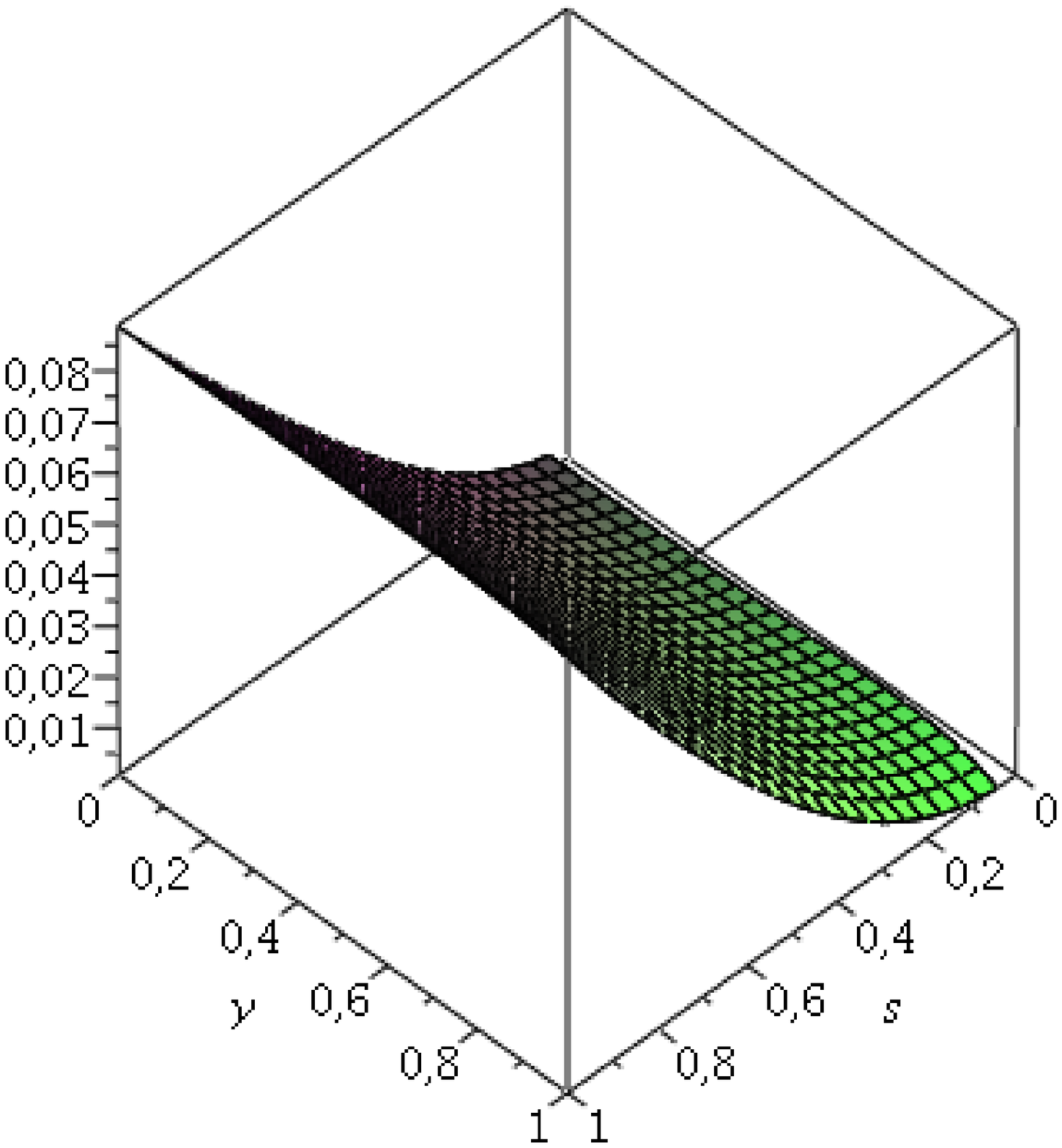}
\caption{\label{t0mt1} {\it  $r_{\Sigma}^2[(- T_0^0)-T_1^1]$ as function of   $y$ and  $s$, with $j=0.1$ and $\tau=2.7$.}}
\end{figure}

Figure 7, shows the component $T_0^3$.
\begin{figure}[h]
 \includegraphics[scale=0.31]{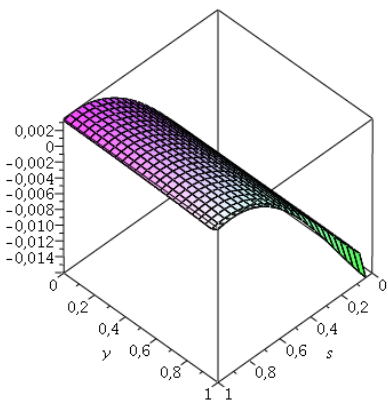}
\caption{\label{t0mt1} {\it  $r_{\Sigma}^3 T_0^3$ as function of   $y$ and  $s$, with $j=0.1$ and $\tau=2.7$.}}
\end{figure}
\section{Discussion}
By extending the general procedure developed in \cite{HHM} to the stationary case, we have been able to build up a physically meaningful source for the Kerr metric, satisfying the matching conditions on the boundary surface of the matter distribution. In spite of the fact that a perfect fluid source for the Kerr metric  might exist \cite{Roos}, our source is necessarily anisotropic in the pressure.

Particular attention deserves the presence of a non--vanishing $T^3_0$ component of the energy--momentum tensor. Indeed, defining as usual an energy--momentum flux vector as: $F^\nu=-V^\mu T_{\nu \mu}$ (where $V^\mu$ denotes the four velocity of the fluid), it appears  that, in the equatorial plane of our system,  energy flows round in circles around the symmetry axis. This result is a  reminiscence of an effect appearing in stationary Einstein--Maxwell systems.  Indeed, in all stationary Einstein--Maxwell systems, there is a non vanishing component of the Poynting vector describing a similar phenomenon \cite{8, 10} (of electromagnetic nature, in this latter case). Thus, the appearance of such a component, seems to be a distinct physical property of rotating fluids, which has been overlooked in previous studies of  these sources.

We have carried out  a systematic search for  the  range of values of $\tau$ and $j$  for which our models exhibit acceptable physical properties. We have focused  on the fulfillment of  Positive Energy Density (P.E.D.) ($-T_0^0>0$), Strong Energy Condition (S.E.C.) ($(-T_0^0)-T_i^i>0$) and Positive Radial Pressure (P.R.P.) ($P_{rr}=g_{11}T_1^1>0$).The results of this  search are  shown in Table I. It appears evident that physically meaningful sources exist for a wide range of values of the parameters.
 
On the other hand, the chosen range of  values of the parameters, incorporates values  considered in the  existing literature, to describe realistic models of  rotating neutron stars and white dwarfs. Let us elaborate on this issue with some detail.

The rotation of the source is determined by the parameters $a, M$ of the exterior solution. Indeed, the rotation parameter $j=J/M^2=a/M$ stands for the dimensionless angular momentum of the rotating source. So, if we restrict ourselves to a sub-extreme Kerr solution ($a<M$), then $j<1$.

 In \cite{laarakers} numerical models of rotating neutron stars are constructed for different EOSs. For each EOS the star's angular momentum, ranges from $J=0$ to the Keplerian limit $J=J_{max}$, and the dependence of the quadrupole moment on the rotation parameter $j$, is established. Within the interval $1.0$ to $1.8$ solar masses for the mass $M$, the parameter $j$ is in the interval  between $0.1$ and  $0.8$.

In \cite{friedman} upper limits on the parameter $j$ are found for representative EOSs; uniformly rotating  neutron star models with maximum mass for various equations of state are studied, and the parameter $j$ does not exceed the value $0.7$. 

In \cite{cook} realistic equations of state for rapidly rotating neutron stars are explored, including a wider range of values for $j$.
 
Finally, we would like to conclude with the following comment. In some static sources it may occur that  $l_z<l_{\rho}$ does not imply that the source is oblate ($q<0$) (see \cite{Bonnor, HHM} for a discussion on this issue).   However, as we have seen, this is not the case of our source.
\begin{widetext}

\begin{table}[htp]
\caption{Fulfillment (F) or violation (V) of different  criteria for good physical behaviour: Positive Energy Density (P.E.D.) ($-T_0^0>0$), Strong Energy Condition (S.E.C.) ($(-T_0^0)-T_i^i>0$) and Positive Radial Pressure (P.R.P.) ($P_{rr}=g_{11}T_1^1>0$). The symbol $^*$ over F means  that although the criterion is fulfilled, nevertheless $\partial_r P_{rr}$ changes its sign in the interval $s \in [0,1]$ within the source.
This table corresponds to an oblate source (due to the rotation) of the Kerr solution for different values of $j$ and a sequence of different values of the parameter $\tau$. }
\begin{center}
P.E.D. / S.E.C. / P.R.P. 
 
\begin{tabular}{ccccccccccccccccc}
\hline\hline
$\tau \diagdown    j$ &$\quad$&
0.9 &$\quad$ &
0.8 &$\quad$ &
0.7 &$\quad$ &
0.5 &$\quad$ &
0.3 &$\quad$ &
0.1 &$\quad$ &
0.05 &$\quad$ &
0.03 
 \vspace{0.2cm} \\  \hline
2.67 &$\quad$ & V V V  &$\quad$ &   V V  V      &$\quad$ & V  V $F^*$   &$\quad$ & V V F &$\quad$ & F  V  F &$\quad$ & F V F &$\quad$ &F V F &$\quad$ &F F F\\ \hline
2.7 &$\quad$ & V V V  &$\quad$ &   V V  V      &$\quad$ & V  V $F^*$   &$\quad$ & V V F &$\quad$ & F  V  F &$\quad$ & F F F &$\quad$ &F F F &$\quad$ &F F F \\ \hline
2.8 &$\quad$ & V V V  &$\quad$ &   V V  $F^*$      &$\quad$ & V  V $F^*$   &$\quad$ & V V F &$\quad$ & F  V  F &$\quad$ & F F F &$\quad$ &F F F &$\quad$ &F F F  \\ \hline
2.9 &$\quad$ & V V V  &$\quad$ &   V V  $F^*$      &$\quad$ & V  V $F^*$   &$\quad$ & F V F &$\quad$ & F  F  F &$\quad$ & F F F &$\quad$ &F F F &$\quad$ &F F F   \\ \hline
3 &$\quad$ & V V $F^*$  &$\quad$ &   V V  $F^*$      &$\quad$ & V  V $F^*$   &$\quad$ & F V F &$\quad$ & F  F  F &$\quad$ & F F F &$\quad$ &F F F &$\quad$ &F F F   \\ \hline
3.1 &$\quad$ & V V $F^*$  &$\quad$ &   V V  $F^*$      &$\quad$ & V  V $F^*$   &$\quad$ & F V F &$\quad$ & F  F  F &$\quad$ & F F F &$\quad$ &F F F &$\quad$ &F F F   \\ \hline
3.5 &$\quad$ & V V $F^*$  &$\quad$ &   F V  $F^*$      &$\quad$ & F  F $F^*$   &$\quad$ & F F F &$\quad$ & F  F  F &$\quad$ & F F F &$\quad$ &F F F &$\quad$ &F F F  \\ \hline
4 &$\quad$ & F F $F^*$  &$\quad$ &   F F  $F^*$      &$\quad$ & F  F F   &$\quad$ & F F F &$\quad$ & F  F  F &$\quad$ & F F F &$\quad$ &F F F &$\quad$ &F F F  \\ \hline
4.5 &$\quad$ & F F $F^*$  &$\quad$ &   F F  F      &$\quad$ & F  F F   &$\quad$ & F F F &$\quad$ & F  F  F &$\quad$ & F F F &$\quad$ &F F F &$\quad$ &F F F   \\ \hline
5 &$\quad$ & F F F  &$\quad$ &   F F  F      &$\quad$ & F  F F   &$\quad$ & F F F &$\quad$ & F  F  F &$\quad$ & F F F &$\quad$ &F F F &$\quad$ &F F F    \\ 
\hline\hline
\end{tabular}
\end{center}
\label{data}
\end{table}

\end{widetext}

\section{Acknowledgments}
This  work  was partially supported by the Spanish  Ministerio de Ciencia e
Innovaci\'on under Research Projects No.  FIS2015-65140-P (MINECO/FEDER), and the Consejer\'\i a
de
Educaci\'on of the Junta de Castilla y Le\'on under the Research Project Grupo
de Excelencia GR234.

\end{document}